\documentstyle[aps]{revtex}
\input epsf

\def\sdtimes{\mathbin{\hbox{\hskip2pt\vrule
height 4.1pt depth -.3pt width .25pt\hskip-2pt$\times$}}}
\begin{document}
\thispagestyle{empty}
\baselineskip 16pt
\parskip 20pt
\rightline{KIAS-P02010}
\rightline{{\tt hep-th}/0204111}
\

\def\sdtimes{\mathbin{\hbox{\hskip2pt\vrule
height 4.1pt depth -.3pt width .25pt\hskip-2pt$\times$}}}

\def\tr{{\rm tr}\,}
\newcommand{\beq}{\begin{equation}}
\newcommand{\eeq}{\end{equation}}
\newcommand{\beqn}{\begin{eqnarray}}
\newcommand{\eeqn}{\end{eqnarray}}
\newcommand{\bde}{{\bf e}}
\newcommand{\balpha}{{\mbox{\boldmath $\alpha$}}}
\newcommand{\bsalpha}{{\mbox{\small\boldmath$alpha$}}}
\newcommand{\bbeta}{{\mbox{\boldmath $\beta$}}}
\newcommand{\blambda}{{\mbox{\boldmath $\lambda$}}}
\newcommand{\bepsilon}{{\mbox{\boldmath $\epsilon$}}}
\newcommand{\ggg}{{\boldmath \gamma}}
\newcommand{\ddd}{{\boldmath \delta}}
\newcommand{\mmm}{{\boldmath \mu}}
\newcommand{\nnn}{{\boldmath \nu}}

\newcommand{\bra}[1]{\langle {#1}|}
\newcommand{\ket}[1]{|{#1}\rangle}
\newcommand{\sn}{{\rm sn}}
\newcommand{\cn}{{\rm cn}}
\newcommand{\dn}{{\rm dn}}
\newcommand{\diag}{{\rm diag}}

\

\vskip 00cm

\centerline{\LARGE\bf First and Second Order  Vortex Dynamics}

\vskip 1.5cm
\centerline{\large\it 
Yoonbai Kim$^{a,}$\footnote{\tt Electronic Mail: yoonbai@skku.ac.kr} and 
Kimyeong Lee$^{b,}$\footnote{\tt Electronic Mail: klee@kias.re.kr} } 
\vskip 3mm
\centerline{$^a$BK21 Physics Research Division and Institute for Basic Science,
Sungkyunkwan University}
\centerline{Suwon 440-746, Korea}
\vskip 3mm

\centerline{$^b$School of Physics, Korea Institute for Advanced Study}
\centerline{207-43, Cheongryangri-Dong, Dongdaemun-Gu, Seoul 130-012, Korea}
\vskip 1.0cm
\vskip 5mm
\begin{quote}
{\baselineskip 16pt 
The low energy dynamics of vortices in selfdual
Abelian Higgs theory is of second order in vortex velocity and
characterized by the moduli space metric. When Chern-Simons term with
small coefficient is added to the theory, we show that a term linear
in vortex velocity appears  and
can be consistently added to the second order expression. We provides
an additional check of the first and second order terms by studying 
the angular momentum in the field theory.   We briefly  explore other
first order  term due to  small background
electric charge density and  also the harmonic potential well for
vortices given by  the moment of inertia. }
\end{quote}

\newpage
\setcounter{footnote}{0}

\section{Introduction}

In the field theory solitons appear as nonperturbative classical
objects which are interacting strongly each other when elementary
quanta of the theory are interacting weakly. Among those solitons
there exists a special class of the so-called selfdual solitons or BPS
solitons. When they can be put together maintaining the BPS property,
there is no static force between them and the solution space of
multi-solitons is uniquely fixed by the moduli parameters.
The description of their dynamics at low energy can be
approximated as the moduli space dynamics. The metric of the moduli
space is induced from the field theory and determines the exact form
of the effective Lagrangian~\cite{manton0}.

This idea has been successfully employed in the low energy dynamics of
BPS magnetic monopoles in Yang-Mills Higgs theories~\cite{atiyah} and
selfdual vortices in Abelian Higgs theories~\cite{samols}. (For the
slow motion of selfdual vortices on Poincare upper half plane, there
exists a beautiful work by Strachan~\cite{strachan}.) There has
been an attempt to find a similar effective Lagrangian for vortices in
pure Chern-Simons Higgs theories~\cite{yoonbai}. While the theory is
intrinsically relativistic, the low energy effective Lagrangian found
in Ref.~\cite{yoonbai} was only of first order in vortex
velocity. This first order effective Lagrangian was shown to be
consistent and describes fractional statistics of vortices.  However
the attempt to find the second order expression was not
successful. (See Ref~\cite{dzia} for other attempts.)

In this paper we start with the selfdual Abelian Higgs system whose low
energy vortex dynamics is well known.  Then we add the Chern-Simons
term with small coefficient as a perturbation to the original theory
and derive the correction to the vortex dynamics, which is the first
order in both vortex velocity and the Chern-Simons coefficient. The
first order expression can then be in the same order as in the second
order one.  By comparing the angular momentum from both field theory
and low energy Lagrangian, we show that the low energy Lagrangian is
consistent.  By using this effective Lagrangian we explore two vortex
dynamics in some detail.  In addition we study briefly the first order
term induced by the small uniform background charge density and also a
small harmonic potential well for vortices given by the quadratic norm
of the Killing vector for the angular momentum, which is the moment of
inertia for rotation.

Our progress is inspired by a recent work in understanding the low
energy dynamics of 1/2 BPS monopoles in N=4 supersymmetric Yang-Mills
theories, when additional Higgs field takes nonzero expectation
value~\cite{dongsu}. The low energy Lagrangian of these monopoles
 consists of the kinetic and the potential parts. The
1/4 BPS or non BPS dyons arise in this effective Lagrangian
naturally. The effective Lagrangian is valid in the regime where both
kinetic and potential energies are much smaller than monopole mass but of
the same order.

Vortices in Abelian Higgs theory was studied in early
seventies~\cite{niel} and shown to be selfdual when the Higgs
coupling has the critical value~\cite{bogo}.  The self-duality can be
found even in the Chern-Simons-Higgs theories~\cite{erick} and the
mixed Maxwell-Chern-Simons theories~\cite{choonkyu}. Vortices in these
theories carry fractional spin and satisfy fractional
statistics~\cite{min,yoonbai}.  Thus the low energy dynamics of these
vortices seems to contain both second order and first order
expressions in velocity.

The low energy effective Lagrangian for vortices in Abelian Higgs
theories is of second order in velocity $v$ and was derived by
Samols~\cite{samols}. When the Chern-Simons term with small
coefficient $\kappa$ is added, we expect that the low energy
Lagrangian should be made of terms of order $v^2$ and terms of $\kappa
v$. Both of these terms are small and seem to be in the same order.
One goal is to derive  terms of order $\kappa v$ in the low energy 
Lagrangian.

Another goal is to show the consistency of the low energy Lagrangian .
We calculate the conserved angular momentum from the field theory and
express it in terms of the moduli space coordinates and velocities. We
show that it is identical to what one gets from the low energy
Lagrangian directly. This is true for both terms of order $v^2$ and 
of order $\kappa v$. 

We should note that the statistics of vortices with nonzero fractional
spin is not given by the naive Aharonov-Bohm phase, rather it is given
by the sum of the Aharonov-Bohm phase and the additional Berry phase
due to the `Magnus force'~\cite{yoonbai}.  This has led to the
explanation for the famous sign flip of vortex spin compared with the
spin of elementary charged particles in the symmetric phase. The sign
flip can be also seen from the transformation $\kappa \rightarrow
-4\pi/\kappa$ of the Chern-Simons coefficient in the mirror map which
makes vortices as elementary charged particles.  Our first order
Lagrangian again captures this interaction between vortices as well.

{}From the study of two vortex dynamics, we can expect nontrivial bound
states with binding energy of order of elementary particle mass when the
Chern-Simons coefficient $\kappa$ is much larger than
one~\cite{yoonbai}. The binding energy would be order of elementary
particle mass. When $\kappa$ has near zero, we do not expect any quantum
bound state. This is consistent with our low energy Lagrangian which
makes sense only for small $\kappa$.

One can ask whether or not there can be other perturbation of the vortex
dynamics by adding other interactions to the theory. One case is to
add a small uniform background electric charge, while keeping the
selfduality.  The first vortex dynamics in this theory is already done
by one of us (K.L.)~\cite{kml}.  We can now add this to the quadratic
Lagrangian when the background charge density $ \rho_e$ is very small
in appropriate unit.  This leads to an additional correction of order
$\rho_e v$ to the low energy Lagrangian.  We just need to translate
the result of Ref.~\cite{kml} to the present notation.  In that work 
the angular momentum in the field theory is shown to be identical to 
that of the low energy effective Lagrangian of vortices. 

More recently selfdual vortices in Abelian Higgs model with a first
order kinetic energy has been studied~\cite{manton2}. The first order
effective Lagrangian for slowly moving vortices has been obtained in
this system and its angular momentum is shown to be consistent with
that of the field theory. In this theory the Galilean symmetry is
preserved and vortices do not carry any intrinsic spin.

One could also add some potential to the vortex dynamics. When the
coupling constant in front of the scalar potential deviates a little
from the critical value, there exists an induced potential between
vortices~\cite{shah}. Unfortunately this induced potential cannot be
expressed in terms of vortex moduli coordinate explicitly unlike the
kinetic term, making the analysis complicated.  One may ask whether
there exists other type of potential to add to the vortex
dynamics. Inspired by the previously mentioned potential term to the
magnetic monopole dynamics~\cite{dongsu,tong}, we add a potential
which is a quadratic norm of the angular momentum Killing vector, that
is, the moment of inertia, arriving at a harmonic potential well for
vortices.

The plan of this paper is as follows. In Sec.~2, we study the selfdual
Abelian Higgs theory with additional Chern-Simons term.  In Sec.~3, we
rederive Samols' result for pure Abelian Higgs theory in somewhat
different gauge. We also show that the angular momentum obtained from
the field theory is identical to that from the Samols' second order
Lagrangian. In Sec.~4, with the Chern-Simons term with small
coefficient, we derive the first order term in the low energy
effective Lagrangian. The angular momentum from the field theory is
again shown to be identical to that from the first order term up to
the constant term. The constant term is related to the intrinsic
vortex spin. In Sec.~5, we study the two vortex interaction in
detail. The anyonic nature of vortices is clearer here.  In Sec.~6,
we conclude with some remarks. In Appendix A we summarize the effect
of the uniform background charge of Ref.~\cite{kml}.  In Appendix B we
explore briefly the harmonic potential well given by the quadratic
norm of the Killing vector of the angular momentum.

\section{Model}

The model we consider is a theory of a complex scalar field $\phi =
fe^{-i\theta}$ coupled to the gauge field $A_\mu$ with the
covariant derivative $D_\mu \phi = \partial_\mu \phi - iA_\mu \phi$.
The Lagrangian density for this model~\cite{choonkyu} is
\beqn
{\cal L} &=&  -\frac{1}{4} F_{\mu\nu} F^{\mu\nu} + \frac{\kappa}{2}
\epsilon^{\mu\nu\rho} A_\mu \partial_\nu A_\rho   + \frac{1}{2}
\partial_\mu N \partial^\mu N  \nonumber \\
& & +  \frac{1}{2} \partial_\mu f \partial^\mu f + \frac{1}{2}
f^2(\partial_\mu \theta + A_\mu)^2    -U(N,f) .
\label{lag}
\eeqn
The kinetic term for the gauge field has the usual Maxwell term and
the parity violating Chern-Simons term. There is a neutral scalar
field $N$ which couples to the matter field. As we are interested in
only the classical aspect of the theory, we put the electric coupling
constant $e=1$ for the convenience. We also scaled the fields and
spacetime and so there is no dimensionful constant in the Lagrangian 
(\ref{lag}).  The Gauss law constraint
for the field configurations is
\beq
\nabla\cdot {\bf E} + f^2(\dot{\theta}+A_0) + \kappa B = 0 ,
\label{gauss0}
\eeq
where $E_i = F_{0i}$ and $B=F_{12}$. 

There is a BPS bound on the energy when the
interacting  potential is chosen to be 
\beq
U= \frac{1}{8}( f^2 -1 - 2\kappa  N)^2 + \frac{1}{2} f^2 N^2 .
\eeq
To see this, let us reexpress the canonical energy density as
\beqn
{\cal H} &=& \frac{1}{2} ({\bf E} \pm \nabla N)^2 + \frac{1}{2} \left\{ B
\mp\frac{1}{2} (1-f^2+ 2\kappa N)\right\}^2  + \frac{1}{2} (\dot{f}^2 + \dot{N}^2) 
\nonumber \\
 & & + \frac{f^2}{2}
(\dot{\theta} + A_0 \mp N)^2 + \frac{1}{2} \left\{\partial_i f \pm
\epsilon_{ij} f (\partial_j \theta + A_j)\right\}^2 \pm \frac{1}{2} B 
\nonumber \\
& &  \pm   N\left\{ \nabla \cdot {\bf E} +  f^2(\dot{\theta}+A_0) +\kappa B 
\right\} .
\eeqn
After the Gauss law is imposed, the energy $H=\int d^2 x {\cal H}$ is
bounded by the total magnetic flux $\Psi = \int d^2 x B $,
\beq
H \ge \frac{1}{2} |\Psi|  .
\label{bps}
\eeq
While the parity is broken by the Chern-Simons term, C and CTP are not
broken and so both vortices and antivortices carry the same mass.

There are two possible vacua: the symmetric phase where $f=0$ and
$N=-1/(2\kappa)$ and the broken  phase where $f=1$ and $N=0$.
Currently we are interested in the symmetry broken vacua.  In the broken
phase there exist topological vortices. For $n$ selfdual vortices 
the phase of the scalar field can be
chosen to be
\beq
\theta = - {\rm Im} \ln \phi = -\sum_{r=1}^n {\rm Arg} ({\bf r}- {\bf
q}_r),
\label{gauge}
\eeq
where the position vectors of vortices are ${\bf q}_r$ with
$r=1,...,n$. From the smoothness condition of the field $\phi$ at the
position of the vortex, the moduli $f$ of $\phi$ should have a simple
zero at each ${\bf q}_r$. The finite energy condition implies that
$\nabla \theta +{\bf A}$ vanishes quickly enough at the spatial
infinity and so the total 
magnetic flux is quantized :
\beq
\Psi = \int d^2 x B  = 2\pi n \; .
\eeq
Antivortices would have opposite winding and so carry negative
magnetic flux. 

The selfdual configurations are those saturating  the energy bound
(\ref{bps}). They satisfy some trivial equations 
\beqn
&&  E+ \nabla N = 0 , \;\;\;  \dot{f}=0, \nonumber \\
&& \dot{N}=  0 , \;\;\;  \dot{\theta} + A_0 - N= 0 .
\eeqn
In the gauge $\dot{\theta}=0$, we get $A_0=N$ and $\dot{A}_i=0$,
implying that the field configuration is static in time. In addition
they satisfy the selfdual equations
\beqn
&& B = \frac{1}{2}(1-f^2+ 2\kappa N),  \label{bbeq1}\\
&& \epsilon_{ij}\partial_i \ln f =  (\partial_j \theta +A_j), \label{bbeq2}\\
&& \nabla^2 N + f^2 N + \kappa B  = 0 , \label{neq1}
\eeqn
where the last one is from the Gauss law (\ref{gauss0}) with $A_0=N$.
The rest mass of the selfdual $n$ vortices is then $n\pi$ and so each
vortex carries mass $\pi$. 

The angular momentum for the selfdual system
can be obtained from the Noether procedure. The angular momentum with
gauge invariant density is 
\beq
J= - \int d^2x  \epsilon_{ij} x^i \left\{ F_{0k}F_{jk} +
\dot{N}\partial_j N +\dot{f}\partial_j f + f^2(\dot{\theta}
+A_0)(\partial_j \theta + A_j) \right\} .
\eeq
We will discuss this quantity for slowly moving vortices in the
following sections.

\section{The Abelian Higgs Theory}

For the well-known $\kappa=0$ case, we review some aspects of selfdual
vortices and their low energy effective Lagrangian obtained by
Samols. This will provide the basic ground for the arguments in coming
sections.  In addition, we give a new nontrivial check of the Samols'
result by showing the the field theoretic angular momentum is 
identical to one from the quadratic Lagrangian of Samols.

The selfdual equations (\ref{bbeq1}) and (\ref{bbeq2}) with
$\kappa=0$ can be put together into a single equation,
\beq
\nabla^2 \ln f^2 + 1- f^2 = 4\pi \sum_r \delta^2 ({\bf r}-{\bf q}_r) .
\label{feq}
\eeq
With the asymptotic value  $f =1$, the solution of the above
equation is uniquely determined by the set $\Gamma_n =\{ {\bf q}_r;
r=1,...,n\}$~\cite{jaffe}. Thus the moduli space of  $n$ selfdual
vortices  is defined by the positions of vortices. Since the
configuration is invariant under the translation $\delta {\bf r} =
\bepsilon$, and $\delta {\bf q}_r = \bepsilon$, we see
that
\beq
\sum_r \frac{ \partial}{\partial {\bf q}_r} \ln f^2 = -
\frac{\partial}{\partial {\bf r}} \ln f^2 .
\eeq
Thus the center of the mass position ${\bf R} = (\sum_r {\bf q}_r) /
n$ constitutes a complex plane $C$. The relative positions of vortices live
on $C^{n-1}$ modulo all permutations of the vortex positions. Calling
the relative moduli space $\tilde{{\cal M}}_n \approx C^{n-1}/S_n$,
where $S_n$ is the permutation group of $n$ objects.  The total moduli
space of $n$ vortices will be
\beq
{\cal M}_n = C \times \tilde{{\cal M}}_n .
\eeq
The energy density of $n$ vortices approaches exponentially quickly to
that of the vacuum  outside vortex centers. As $\int d^2x (1-f^2) =
2\pi n$ for $n$ selfdual vortices, one can regard each vortex as a
particle of area $2\pi$ and incompressible. While there is no
repulsive force between vortices, $n$ vortices occupy $2\pi n$ area.

It is convenient also  to use the complex coordinates 
\beq
 z = x^1+ ix^2, \;\; 
 z_r = q^1_r + iq^2_r \;\; . 
\eeq 
Note that $\partial_z = (\partial_1 - i \partial_2)/2$.  By expanding
around a vortex position $z_r$ of the solution of Eq.~(\ref{feq}), we
get
\beqn
\ln f^2 &=& \ln |z-z_r|^2 + a_r + \frac{1}{2} \left\{ b_r (z-z_r) +
\bar{b}_r (\bar{z}-\bar{z}_r)\right\} \nonumber \\
&&  + c_r (z-z_r)^2 + \bar{c}_r
(\bar{z}-\bar{z}_r)^2 - \frac{1}{4} |z-z_r|^2 + {\cal O}(|z-z_r|^3) .
\label{fexp}
\eeqn
Since the field 
\beq
\Phi = \ln \frac{f^2}{\prod_r |z-z_r|^2} 
\eeq
is nonsingular everywhere, we can see that 
\beq
 b_r = \sum_{s\ne r} \frac{2}{z_r-z_s} + \tilde{b}_r
\label{beqn}
\eeq
with nonsingular function $\tilde{b}_r $.  Note that $b_r$ vanishes
exponentially when vortices are separated from each other. The
functions $b_r(z_s,\bar{z}_s)$'s will play the crucial role in the
following argument.

\subsection{Samols' Result}

We are interested in the low energy dynamics of these selfdual
vortices. As the moduli coordinates $z_r$'s characterize the
configuration uniquely and describe the zero modes,  the time
evolution of the field configuration will be approximated by the time
evolution of the moduli coordinates $z_r(t)$. We choose the gauge
where
\beq
\theta({\bf r}, t) = - \sum_r {\rm Arg} ({\bf r} - {\bf q}_r(t))
\label{gauge1}
\eeq
through the time evolution. (Samols has chosen the Weyl gauge $A_0=0$ and
so his $\theta(t)$ is more complicated.)

To obtain the effective Lagrangian for the $z_r(t)$ variables, let us
calculate the field theoretic Lagrangian for given initial data
which are made of the field `position' and `velocity'. The field
`position' would be the self field configuration, ${\bf A}({\bf r};{\bf
q}_r)$ and $\phi({\bf r};{\bf q}_r)$.  The field `velocity' is composed of
$E_i$ and  $D_0 \phi$ which could be regarded also as field `momentum'.  
The time
derivatives of the fields are given by $\dot{f} = \sum_r \dot{\bf q}_r
\cdot \partial f /\partial {\bf q}_r$, and so on.  Once we choose the
above gauge (\ref{gauge1}), we cannot require $A_0=0$ anymore. The
initial data should satisfy the Gauss law constraint which fixes the
initial $A_0$ by 
\beq
\partial_i E_i  + f^2 (\dot{\theta}   + A_0)   = 0 .
\eeq
This can put into a form 
\beqn
\partial_i \partial_0 (\partial_i \theta + A_i)
- \partial_i^2 (\dot{\theta} + A_0 ) + f^2(\dot{\theta}+A_0) &=&
\partial_i[\partial_0,\partial_i]\theta  \nonumber\\
& =& 2\pi \sum_r 
\dot{{\bf q}}_r\times \frac{\partial}{\partial {\bf q}_r} \delta^2({\bf
r}-{\bf q}_r) .
\eeqn
Due to the selfdual equation (\ref{bbeq2}), the  first term vanishes and 
the above equation becomes
\beq
\nabla^2 {\rm Im}\; \eta - f^2 {\rm Im}\; \eta = - 2\pi \sum_r 
\dot{{\bf q}}_r \times \epsilon_{ij}
 \frac{\partial}{\partial {\bf q}_r}
\delta^2({\bf r}-{\bf q}_r), 
\eeq
where we have introduced a new field, 
\beq
\eta = \frac{ D_0 \phi}{\phi} = \partial_0 \ln f -i(\dot{\theta}+ A_0).
\label{etafeq}
\eeq
{}From
Eq.~(\ref{feq}) for $f$, we can also get the similar equation for
${\rm Re} \; \eta$. Together we get the equation for $\eta$, 
\beq
4\partial_z \partial_{\bar{z}} \eta - f^2 \eta = 4\pi \sum_r \dot{z}_r
\partial_z \delta^2(z-z_r) ,
\label{etaeq}
\eeq
where $\delta^2(z-z_r) = \delta^2({\bf r}-{\bf q}_r)$.  With the
boundary condition $ \eta=0$ at spatial infinity, the unique solution
of this equation (\ref{etaeq}) is
\beq
\eta = \sum_r \dot{z}_r\frac{\partial }{\partial z_r}\ln f^2 .
\label{etasol}
\eeq
%
%

Now we have the initial field data satisfying the Gauss law. 
Their field theoretic  Lagrangian becomes
\beq
L =  - n \pi +\frac{1}{2} \int_C d^2x \left\{ |E_1 +iE_2|^2 +
|D_0\phi|^2 \right\} ,
\eeq 
where $n\pi$ is the rest mass of $n$ vortices.  As the initial field
configuration is smooth everywhere, we can take out the positions of
vortices from the integration region without changing the value of
the integration. We call this region $\tilde{C}= C-\Gamma_n $.  On
this region $\tilde{C}$ we see that
\beq
E_1 + i E_2 = -2i\partial_{\bar{z}} \eta .
\label{eetaeq}
\eeq
On $C$, the above relation does not hold as there are delta functions
which do not vanish on $\Gamma_n$, but they are not in our
integration region anymore. We will use this technique devised by
Samols again and again with profitable results. From the $\eta$ field,
we also get the initial field `velocity', $D_0\phi $.  Thus the low
energy effective Lagrangian of order $v^2$ becomes
\beqn
L_{v^2}= L +n \pi &=& \frac{1}{2} \int_{\tilde{C}} d^2x \left\{ 4
|\partial_{\bar{z}}\eta|^2 + f^2 
|\eta|^2 \right\} \nonumber  \\ 
&=& 2\int_{\Gamma_n} \partial_z   (\bar{\eta}\partial_{\bar{z}} \eta)
\nonumber \\
&=& -\sum_r \oint_{q_r} d\bar{z}\; \bar{\eta}\partial_{\bar{z}}\eta ,
\label{v2s}
\eeqn
where we have used Eq.~(\ref{etaeq}).

Near a vortex position $z_r$,  the expansion (\ref{fexp}) leads to 
\beq
\eta =  - \frac{\dot{z}_r}{z-z_r} + {\rm smooth}\;\; {\rm terms},
\label{etas}
\eeq
and 
\beq
\partial_{\bar{z}} \eta = \frac{1}{4} \dot{z}_r + \sum_s \dot{z}_s
\frac{\partial \bar{b}_r}{\partial z_s} + {\cal O}(|z-z_r|).
\eeq
After boundary  integrations, we get 
\beq
L_{v^2} =  \frac{\pi}{2} g_{rs} \dot{z}_r \dot{\bar{z}}_s ,
\label{samols}
\eeq
where the moduli space metric is
\beq
g_{rs} = \delta_{rs} + 2 \frac{\partial \bar{b}_s}{\partial z_r} .
\eeq
Since the $L_{v^2}$ is real, 
\beq
\partial \bar{b}_s/\partial z_r = \partial b_r/\partial \bar{z}_s.
\label{bzeq}
\eeq
When vortices are coming close to each other, the $b_r$'s behave as
in Eq.~(\ref{beqn}), and so the metric is regular. The kinetic energy
is of order $v^2$ quantity and is correct when $v\ll 1$. The corrections
of order $v^4$ are negligible in this nonrelativistic or low energy
limit. The moduli space is K\"ahler as the two form
\beq
w = \frac{i}{4} g_{rs} dz_r d\bar{z}_s
\eeq
is closed due to Eq.~(\ref{bzeq}).  Our complex coordinates $z_r$'s are
holomorphic coordinates with respect to this K\"ahler form. 

\subsection{Angular Momentum}

The above effective Lagrangian has conserved angular momentum. We
calculate it for the field theoretic initial `data' of vortices
moving slowly. Similar to the Lagrangian, we can express the field
theoretic angular momentum in terms of moduli parameters, which turns
out to agree with the one obtained from the low energy effective
Lagrangian.

The field theoretic  conserved angular momentum is 
\beq
J = -\int_C d^2x \; \epsilon_{ij} x^i \left\{ F_{0k} F_{jk} +
\dot{N}\partial_j N + 
\dot{f}\partial_j f + f^2 (\dot{\theta} + A_0)(\partial_j \theta
+ A_j)\right\} .
\label{angular1}
\eeq
For the given initial data, the angular momentum becomes
\beq
J =  -2i  \int_{C-\Gamma_n} d^2x \left\{  z
\partial_z(\bar{\eta}\partial_z\partial_{\bar{z}} \ln f^2 )- 
\bar{z}\partial_{\bar{z}}  \eta \partial_z\partial_{\bar{z}}  \ln
f^2 \right\} .
\eeq
Now we note that, on $C-\Gamma_n$, 
\beq
z\partial_z (\bar{\eta}\partial_z\partial_{\bar{z}} \ln f^2 )
= \partial_{\bar{z}}(z\partial_z\bar{\eta}\partial_z \ln f^2)
\eeq
due to the equations (\ref{feq}) and (\ref{etaeq}) satisfied by $f^2$
and $\eta$.  Thus, the integration can be converted to the boundary
integrations,
\beq
J = \sum_r \left( 
\oint_{z_r} dz z \; \partial_z \bar{\eta} \;\partial_z \ln f^2 + c.c.\right).
\eeq
{}From the expansions of $\ln f^2$ and $\eta$  in  Eqs. (\ref{fexp})
and (\ref{etas}) around $z=z_r$, we get
\beq
J = \frac{\pi i}{2} g_{rs}(z_r\dot{\bar{z}}_s- \dot{z}_r \bar{z}_s) 
\label{angv2}
\eeq
which is exactly what we would get from the reduced Lagrangian
(\ref{samols}). This is one more consistency check on  the low
energy effective Lagrangian (\ref{samols}).

Under the spatial translation $x^i \rightarrow x^i + a^i$, the angular
momentum gets shifted by 
\beq
J \rightarrow J + \epsilon_{ij} a^i P^j .
\eeq
This is true both in the field theory and in the particle dynamics,
and so does not provide any additional check.
The  field theoretic linear momentum is 
\beq
P^i = - \int d^2x \left\{\epsilon_{ij} E_j B + \dot{N} \partial_i N +
\dot{f}\partial_i 
f + f^2(\dot{\theta} +A_0)(\partial_i \theta + A_i) \right\} .
\eeq
For the selfdual configurations, the complex momentum $P=P^1+iP^2$  becomes
\beqn
P &=&  \int_{C - \Gamma_n} d^2 z \,\partial_{\bar{z}} \left\{
 \eta(1-f^2) \right\} \\ 
&=& \pi \sum_r \dot{z}_r 
\eeqn
which is consistent with one from the Newtonian dynamics 
due to the translation invariance, $\sum_r \partial \bar{b}_s
/\partial z_r = 0$.

\section{The Chern-Simons Term}

When $\kappa\neq 0$, we go back to the theory in Sec.~2. For
the BPS configuration in the gauge (\ref{gauge}), all fields are
static in time and so $A_0=N$.  Especially the BPS equations 
(\ref{bbeq1})-(\ref{neq1}) for
vortices in the broken phase, $f=1$ and $N=0$, become two coupled equations
\beqn
&& -\nabla^2 N + f^2 N + \frac{\kappa}{2}(1-f^2+2\kappa N) = 0 ,\\
&& -\nabla^2 \ln f^2 + 1-f^2 - 2\kappa N = 4\pi \sum_r \delta^2
(z-z_r) .
\eeqn
The zero modes of the vortex configuration are again given by the set of
the positions of selfdual topological vortices~\cite{LMR}.
The vortex configuration exists in the broken phase~\cite{dongho}, and
is expected to be unique for given vortex position.
For $\kappa=0$, the above equations become Eq.~(\ref{feq})
for $f^2$, which studied in the previous section, and a homogeneous
equation for $N$, which has a trivial solution $N=0$. For small
$\kappa$, the leading correction is $N$ field of order $\kappa$. The
higher order corrections to $f^2$ is of order $\kappa^2$ and that to
$N$ is of order $\kappa^3$.  Here we consider only the leading order
$\kappa$ correction.  The equation for the $N$ field to order $\kappa$
is
\begin{equation}
 -\nabla^2 N + f^2 N + \frac{\kappa}{2} (1-f^2) = 0 ,
\label{neq}
\end{equation}
where $f^2$ satisfies Eq.~(\ref{feq}). We can find the closed  form for $N$
\begin{eqnarray}
N &=& \frac{\kappa}{4} \left\{   \sum_r ({\bf r} -{\bf q}_a)\cdot
\frac{\partial}{\partial {\bf q}_r} \ln f^2 \right\}  \nonumber \\
&=&\frac{\kappa}{4} \left\{ -2n+  \sum_r ({\bf r} -{\bf q}_a)\cdot
\frac{\partial}{\partial {\bf q}_r} \ln \frac{f^2}{\prod_b|{\bf r}-{\bf
q}_b|^2 } \right\} .
\label{selfdual2}
\end{eqnarray}
(This solution is inspired by the case in Ref.~\cite{kml} where the Gauss law
was solved similarly.) The $N$ field is smooth everywhere and
vanishes at spatial infinity.

A BPS configuration made of $f$ and $N$ fields is static but has
nontrivial field `momentum' of order $\kappa$ as $E_i = - \partial_i
N$ and $\dot\theta + A_0 = N$. Thus selfdual vortices carry nonzero
electric charge
\beqn
Q &=& \int d^2 x f^2 (\dot\theta + A_0) \nonumber \\ 
 &=& 2\pi n   \kappa
\eeqn
which is exact in all order in $\kappa$. The static selfdual
configuration also carries a nonzero  angular 
momentum. For a rotationally symmetric configuration of $n$ vorticity,
the exact angular momentum is
\beqn
J_\kappa &=& -2\pi \int_0^\infty dr \frac{d}{dr}  \left\{ rN'\bar{A} -
\frac{\kappa}{2} \bar{A}^2 \right\} \\
&=& - \pi \kappa n^2 ,
\eeqn
where we used $\partial_i \theta + A_i = \bar{A}(r)
\hat{\varphi}^i/r$.  Thus, we see that a single vortex carries charge
$2\pi\kappa$ and spin $s=-\pi \kappa$.

When $n$ vortices are on top of each other, the total angular momentum
is $n^2 s$. On the other hand, the total angular momentum becomes just the 
sum $n s$ when $n$ vortices are in large separation. 
This looks strange but is
expected from the spin statistics as we have explained in
Ref.~\cite{yoonbai}.  We review this briefly in the section where two
vortex dynamics is discussed. We will argue that our effective
Lagrangian is valid when $\kappa$ is very small.

\subsection{Effective  Lagrangian}

The low energy effective Lagrangian can be obtained again by
evaluating the field theoretic Lagrangian for given field initial
data which are made of the field `position' and field `velocity'.
The leading order of the `position' of the field should be given by $f$ and
$N$.  For the BPS configurations, there exist still some nontrivial
initial field momenta of order $\kappa$.  We now want to include the
field momenta of order $v$.

The initial field `momentum' is divided into quantities of order
$\kappa$ and those of order $v$. The order $\kappa$ terms are just
from the $N$ field. The additional `momentum' would be proportional to
$v$ and given by the previous section. Basically, the initial
`momentum' of the vector field and scalar fields change from
Eq.~(\ref{eetaeq}) and  Eq.~(\ref{etafeq}) to  
\beqn
&& E_1+  iE_2 + ( \partial_1 N + i\partial_2 N) = -2i \partial_{\bar{z}}
\eta ,\\
&&  \partial_0 f -i(\dot{\theta}+A_0-N) =   \eta  ,
\eeqn
on $C-\Gamma_n$ with $\eta$ in Eq.~(\ref{etasol}).

We use this initial field data to evaluate the field
theoretic Lagrangian. The leading zeroth order is the minus of the
rest mass.  The next order terms are made of terms of order $v^2,
\kappa v$, and $\kappa^2$. We can see that order $\kappa^2$ should
vanish as it represents just a shift of the rest mass or an
additional potential, which is not there. After direct calculation of
the Lagrangian, we get 
\beq L = -\pi n + L_{v^2} + L_{\kappa v1} +
L_{\kappa v2} ,
\eeq
where
\beq
L_{v^2} = \frac{1}{2} \int d^2 x   \left\{  ({\bf E}- \nabla N)^2 +
\dot{f}^2 + f^2 (\dot{\theta}+ A_0 - N)^2 \right\},
\eeq
\beq
L_{\kappa v1} = \kappa \int d^2 x \left\{  
\frac{1}{2}(\dot{A}_1 A_2 - \dot{A}_2 A_1) -  B \dot{\theta} \right\} ,
\eeq
and
\beq
L_{\kappa v2} = \int d^2 x \left\{ -({\bf E}+\nabla N)\cdot \nabla N + 
f^2  (\dot{\theta}+ A_0- N) N + \kappa B(\dot{\theta} + A_0-N) \right\} .
\eeq 
The $v^2$ order Lagrangian  $L_{v^2}$ comes from the fact that
vortices have the initial velocity. As there is no interference of 
order $\kappa$ terms, $L_{v^2}$ is identical to the Samols' result (\ref{v2s})
discussed in the previous section.

Among the terms in $L_{kv1}$, it is easy to see the contribution from
$A_2\dot{A}_1 - A_1\dot{A}_2$ vanishes due to the selfdual condition.
The Lagrangian $L_{\kappa v1}$ is exactly identical with the Lagrangian we
have obtained for the slow motion of vortices in the pure Chern-Simons
Higgs theory in Ref.~\cite{yoonbai}.  Even though the $f$ field in
that theory obeys a different equation, the character of the low
energy Lagrangian is identical. While $\dot{\theta}$ is singular 
at each vortex position, it is not delta-function singularity. On
$C-\Gamma_n$, we see that
\beq
\dot{\theta} B = \sum_r \dot{q}_r^i \partial_j \left\{
( \delta_{il}\delta_{jk} + \delta_{ik}\delta_{jl}
-\delta_{ij}\delta_{kl} ) A_k \partial_l  \ln |{\bf r}-{\bf q}_r|
\right\}.
\eeq
Thus
\beq
L_{\kappa v1} = 2\pi \kappa \sum_r \dot{{\bf q}}_r \cdot  {\bf A}
( {\bf r}) |_{{\bf r} = {\bf q}_r } .
\label{kappav1}
\eeq
After inserting the asymptotic form for $\ln f^2$ in Eq.~(\ref{fexp}), 
we get the Lagrangian
\beq
L_{\kappa v 1} = \frac{ \pi \kappa}{2} \sum_r i(\dot{z}_r \tilde{b}_r -
\dot{\bar{z}}_r \bar{\tilde{b}}_r ) .
\label{kapv1}
\eeq
This defines a  natural one form $\Omega_1$ on the moduli space  
\beq
\Omega_1 = \frac{\pi\kappa }{2} \sum_r i ( \tilde{b}_r dz_r - 
\bar{\tilde{b}_r} d\bar{z}_r) . 
\eeq
{}From Eq.~(\ref{beqn}), we see that $\tilde{b}_r$ is smooth when
vortices are coming together but has a long range tail when they are
separated.  This is what we need for the statistical phase interaction
between vortex anyons.

One can impose the Gauss law on  $L_{\kappa v 2}$ by many different
ways. The natural one is to replace $f^2 N$ by using Eq.~(\ref{neq})
for $N$, which leads to 
\beqn
&& -({\bf E}+\nabla N)\cdot \nabla N + f^2 N(\dot{\theta}+A_0-N) + 
\kappa B(\dot{\theta}+A_0-N) \nonumber
\\ && = \partial_i \{\partial_i N (\dot{\theta}+
A_0 -N) \} - \partial_i \{N\partial_0 (\partial_i \theta + A_i)\}
\eeqn
on $C - \Gamma_n$.  After contour integration, we get 
\beq
L_{\kappa v 2} =  \pi \sum_r  \dot{{\bf q}}_r \times \nabla N
|_{{\bf r}={\bf q}_r} .
\label{kappav2}
\eeq
We can find the asymptotic formula for the $N$ field and so we get 
\beq
L_{\kappa v 2} = \frac{\pi \kappa  }{8}  \sum_r \left( 
i \dot{z}_r H_r + c.c. \right) ,
\label{kapv2}
\eeq
where
\beq
H_r= -b_r + \sum_{s\ne r} (z_r-z_s) \frac{\partial
b_r}{\partial z_s} + \sum_{s\ne r} (\bar{z}_r-\bar{z}_s
)\frac{\partial b_r}{\partial \bar{z}_s}\; .
\label{hreq}
\eeq
This Lagrangian is not changed when we replace $b_r$ by the regular
$\tilde{b}_r$. Thus, this Lagrangian is regular everywhere and has no
long range tail. This means that this Lagrangian does not provide any
nontrivial statistics between vortices. This Lagrangian is purely
local and affects only the short distance interaction between
vortices.  This Lagrangian introduces to the moduli space another one
form,
\beq 
\Omega_2 =\frac{\pi\kappa }{8} \sum_r \left(  i dz_r  H_r +
c.c. \right) .
 \eeq
As there is no long range tail for this form, the field strength
$d\Omega_2$  over any two dimensional plane will have zero net magnetic flux.

The sum of the Lagrangians in Eq.~(\ref{kapv1}) and Eq.~(\ref{kapv2}) forms
the order $\kappa v$ effective Lagrangian ${\cal L}_{\kappa v}$ such as
\beq
{\cal L}_{\kappa v} = {\cal L}_{\kappa v 1} + {\cal L}_{\kappa v 2} .
\eeq
When both $\kappa $ and $v$ are small and of the same order, it can be 
regarded as the
same order Lagrangian ${\cal L}_v^2$ of Samols (\ref{samols}).

\subsection{Angular Momentum}

For the given initial field data, we can calculate the conserved
angular momentum (\ref{angular1}).  The term $\dot{N}\partial_j N$ is
of order $ \kappa^2 v$ and can be neglected in the order we are
working on. The angular momentum can be  decomposed into terms of
order $\kappa$ and those of order $v$. The order $v$ terms are
\beq
J_{v} = -\int d^2x \; \epsilon_{ij}x^i \left\{ \epsilon_{jk}
(E_k+\partial_j N) B + 
 \dot{f}\partial_j f+  f^2
(\dot{\theta} + A_0-N) (\partial_j \theta + A_j) \right\} .
\eeq
For a given  initial data, this expression is identical to
Eq.~(\ref{angv2}) as expected.

The order $\kappa$ correction to the angular momentum can be
expressed
\beq
J_\kappa = -\int d^2x \epsilon_{ij} x^i \left\{ f^2 N (\partial_j
\theta + A_j) - \epsilon_{jk} \partial_k N B \right\} .
\eeq 
The order $\kappa$ terms indicate the intrinsic angular momentum of
the selfdual configuration.  Using Eq.~(\ref{neq1}) for $N$, we can
divide  $J_\kappa$ as a sum of 
\beq
J_{\kappa 1} = \kappa \int d^2 x \; \epsilon_{ij} x^i B (\partial_j
\theta +A_j)
\eeq
and 
\beq
J_{\kappa 2} = -\int d^2 x \; \epsilon_{ij} x^i \left\{ (\partial_j
\theta +A_j) \partial_k^2 N - \epsilon_{jk} \partial_k N B \right\}.
\eeq

If we define the singular vector potential $C_i$ as $C_i = \partial_i \theta +
A_i$, the first angular momentum $J_{\kappa 1}$ is the expression we
have obtained in Ref.~\cite{yoonbai} for the pure Chern-Simons Higgs theory.
Employing the method in Ref.~\cite{yoonbai}, we have 
\beqn
J_{\kappa 1} &=& \kappa \int d^2x \partial_i \left( \frac{x^i}{2} C_j^2
- C_i x^j C_j \right) \nonumber \\
&=& - \pi \kappa n^2 + 2\pi \kappa \sum_r {\bf q}_r \times
 {\bf A} |_{{\bf r}={\bf q}_r} .
\label{jkv1}
\eeqn
By the similar manipulation,  the second angular momentum becomes 
\beqn
J_{\kappa 2} &=& \int_{C-\Gamma_n} d^2x \; \partial_i \left( 
\epsilon_{jk} x^j \partial_i C_k - \epsilon_{jk}x^j N \partial_k C_i 
- \epsilon_{ij} NC_j \right) \nonumber \\
&=& -\pi \sum_r {\bf q}_r \cdot \nabla  N|_{{\bf r}={\bf q}_r} .
\eeqn
We can see that this field theoretic angular momentum of order
$\kappa$ is identical to the angular momentum implied by the low
energy effective Lagrangian of order $\kappa v$ given in
Eq.~(\ref{kappav1}) and Eq.~(\ref{kappav2}) up to a constant term $-\pi
\kappa n^2$ of Eq.~(\ref{jkv1}).

\section{Two Vortex Dynamics}

\subsection{Old Results}

The moduli space ${\cal M}_2$ of two vortices can be split into $R^2$
for the center of mass motion and $\tilde{{\cal M}}_2$ for the
relative motion. We introduce the center of mass position $Z=
(z_1+z_2)/2$ and the relative position $\zeta = (z_1-z_2)/2$. In the
center of the mass frame, there is obvious symmetry under the exchange
of two vortices, which implies
\beq
b_2(\zeta) = - b_1(\zeta)= b_1(-\zeta) .
\eeq
The  moduli space metric
becomes
\beq
L_{\rm two}  = \pi |\dot{Z}|^2 + \pi\biggl( 1 + 
\frac{\partial \bar{b}_1}{\partial \zeta}\biggr) |\dot{\zeta}|^2 .
\eeq
As shown in Ref~\cite{samols}, in the polar coordinate $\zeta = \sigma
e^{i\varphi}$,  $b_1(\zeta) = b(\sigma)e^{-i\varphi}$ and
so the metric for the relative motion becomes
\beq
L_{v^2} = \pi  F^2(\sigma)(\dot{\sigma}^2 + \sigma^2 \dot{\varphi}^2) 
\label{twometric}
\eeq
with $0\le \varphi < \pi$ and
\beq
F^2(\sigma) = 1  + \frac{1}{\sigma}(\sigma b)'\;\; .
\eeq
{}From Eq.~(\ref{twometric}) and the fact that the metric is smooth at
the origin in terms of the well defined coordinate $\omega=\sigma^2
e^{2i\varphi}$, one can get the asymptotic expansion as follows:
\beq
b(\sigma) = \frac{1}{\sigma} - \frac{\sigma}{2} +  
c \sigma^3 + {\cal O}(\sigma^4) 
\eeq
near $\sigma =0 $ with nonzero positive $c$. For large $\sigma$, we
get $ b(\sigma) = {\cal O}(e^{-2\sigma}) $.  This leads to an integral
formula $\int_0^\infty d\sigma \sigma (1-F^2(\sigma)) = 1$.  The
K\"ahler potential for the metric is $K(\sigma) = \sigma^2/4 +
\int^\sigma d\sigma' b(\sigma')$. 
The conserved angular momentum for
the relative motion is
\beq
J_{v} = 2\pi F^2(\sigma) \sigma^2 \dot{\varphi}\; . 
\eeq 

The motion of vortices on moduli space is given by the geodesic. The
relative moduli space of two vortices is a cone of deficit angle 180
degrees. Thus, the head-on collision of two vortices leads to 90
degree scattering~\cite{ruback}.  The scattering of two vortices in
the field theory was explored in detail numerically~\cite{rebbi} and
was found agreeing with the results from the effective
Lagrangian~\cite{samols}.

\subsection{First Order  Term }

As shown in the previous section, there are two kinds of contribution up to the
first order terms in the low energy Lagrangian. The relative part of
the first contribution for two vortices becomes
\beq
L_{\kappa v1 }=  -2\pi \kappa (\sigma b-1) \dot{\varphi} \; .
\eeq
This first order Lagrangian is smooth at $\sigma=0$ and has a long
range tail at the large distance.  Its corresponding conserved angular
momentum (\ref{jkv1}) becomes 
\beqn
J_{\kappa 1}(\sigma) &=&  -4\pi \kappa  -2\pi \kappa (\sigma b -1)  
\nonumber \\ 
&=& -2\pi \kappa - 2\pi \kappa \sigma b
\eeqn
which interpolates from $-4\pi\kappa$ at the origin to $-2\pi\kappa$
at  large distance. 

The second contribution becomes 
\beq
L_{\kappa v 2} = \frac{\pi \kappa}{2} \sigma (\sigma b)' \dot{\varphi}.
\eeq
This first order Lagrangian is smooth at the origin and vanishes
exponentially at infinity. There is no nontrivial long range
interaction induced by this Lagrangian. The corresponding conserved
angular momentum is
\beq
J_{\kappa 2} = \frac{\pi\kappa}{2}\sigma (\sigma b)' .
\eeq

The low energy effective Lagrangian for two vortices in the center of
mass frame is then 
\beq
L_{\rm two} = \pi F^2(\sigma) (\dot{\sigma}^2 +
\sigma^2\dot{\varphi}^2)  + {\cal A}(\sigma) \dot{\varphi},
\eeq
where 
\beq
{\cal A}(\sigma) = 2s (\sigma b-1) -
 \frac{s}{2} \sigma (\sigma b)' 
\eeq
with vortex spin $s=-\kappa \pi$. The conserved Noether charge is
\beq
M = 2\pi\sigma^2 F^2 \dot{\varphi} + {\cal A},
\eeq
while the total angular momentum is $J_{\rm tot}= 4s + M$.  The  conserved
Hamiltonian is 
\beq
{\cal H} = \frac{1}{4\pi F^2}\left\{ P_\sigma^2 +  \frac{(M-{\cal
A})^2}{\sigma^2} \right\} .
\label{hamiltonian}
\eeq 
Note that ${\cal A}$ has a long range interaction even though it is
smooth at the origin.

\subsection{Quantum Mechanics}

Let us start with recalling a general feature of two anyon dynamics in
the center of mass frame. With fractional spin $s$, the orbital
angular momentum is $2s+ 2l$ and so the total angular momentum is $4s+
2l$. While anyon and antianyon carry the same sign spin $s$,  their
orbital angular momentum eigenvalue is $-2s +2l$ and so they can
annihilate each other to the vacuum.

Anyons can be treated as bosons with electric charge and magnetic
flux, and so have a long range Aharonov-Bohm like
interaction. Equivalently, they can be treated as particles of
fractional spin without any further long range gauge interaction.
This can be done also for the quantum mechanics of two Chern-Simons vortices.

Let us consider the quantum mechanics of this system briefly.
The Hamiltonian (\ref{hamiltonian}) for the relative motion in the 
operator becomes
\beq
{\cal H} =  -\frac{1} {4\pi \sigma F^2} \partial_\sigma(\sigma 
\partial_\sigma
) + \frac{1}{4\pi \sigma^2 F^2} (-i\partial_\varphi - {\cal A})^2 .
\eeq
After quantization, we treat vortices as identical bosons with a long
range magnetic interaction. Then the wave function $\Phi_{\rm boson}$
for the two bodies should be single-valued under exchange.  As the range of
$\varphi$ is $[0,\pi]$, $\Phi_{\rm boson} \sim e^{i2l \varphi}$ and
the orbital angular momentum $M= -i\partial/\partial \varphi  $
has the even integer eigenvalues $2l$'s.  The wave function is well defined on 
the moduli space. 

However, this is not the whole story. The eigenvalue of the conserved
total angular momentum becomes
\beq
J_{\rm tot} =  4s + M = 4s+ 2l 
\eeq
of which $2s$ would be  the sum of the vortex intrinsic spin.
Thus, it is better to interpret the $2l+2s$ as the eigenvalue of the
new orbital angular momentum $M_{orb} = M +2s$.
 Then the anyonic wave  function
\beq
\Phi_{\rm anyon} = e^{2is \varphi} \Psi_{\rm boson}
\eeq
is the eigenstate of the $M_{\rm orb} = -i \partial_{\varphi}$ with
eigenvalue $2l+2s$ as expected for two anyons of spin $s$. The
Hamiltonian for $\Psi_{\rm anyon}$ becomes 
\beq
{\cal H}_{\rm anyon} = - \frac{1}{4\pi \sigma F^2} \partial_\sigma (\sigma
\partial_\sigma) + \frac{1}{4\pi \sigma^2 F^2 } (-i\partial_\varphi - {\cal A}')^2
\eeq
with  ${\cal A}'= 2s\sigma b -s\sigma(\sigma b)'/2$, which is smooth
and provides a short range interaction. This is exactly what is
expected for anyons.

Classically, the orbital angular momentum of two selfdual vortices
changes from $2s$ to zero as their separation increases.  For nonzero
$s$, two overlapped vortices cannot escape to infinity with very small
energy as argued in Ref.~\cite{yoonbai}.  The reason is that their
interaction is short ranged and the spatial motion cannot carry enough 
orbital angular momentum when the available kinetic energy is
arbitrarily small as one needs a large impact parameter. Similarly, two
vortices in large separation cannot come together with very low initial
kinetic energy. 

To see this quantum mechanically, we first note that the classical
approximation is good when the classical quantity is much larger than
$\hbar$. Thus the classical picture of vortex spin is fine if the
vortex spin $s=-\pi\kappa $ is much larger than unity. Assuming the
quantum correction is very small, say due to the underlying
supersymmetry, one can see that the selfdual two vortex states,
saturating the BPS energy, could have the orbital angular momentum $2s
+ 2l$ with the integer $l$ which ranges from zero to $\approx -s$, depending
on how close they are. From the classical picture, two vortices will
be on top of each other when $M_{\rm orb}= 2s$ and will be far apart
when $M_{\rm orb} = 0$. As the interaction is short ranged, the
quantum states with $|M_{\rm orb}| =| 2s+2l| \ge 1 $ would be bound
states of vortices of zero energy with size of vortex core. The bound
energy of these states without supersymmetry would be the order of
elementary particle mass. The quantum state with $M_{\rm orb} = 2s+
2l\approx 0$ would be the ground state of states of the continuum
energy.

We have argued that the moduli space approximation is good when 
$s=-\pi\kappa$ is very small. Quantum mechanically we do not
expect any bound state and states with large bound
energy. Thus, our approximation of the low energy dynamics of vortices
in small $v$ and $\kappa$ limit with Lagrangians of order $v^2$ and
$\kappa v$ is consistent.

\section{Conclusion}

{}From the field theoretic Lagrangian, we have obtained the low energy
effective Lagrangian for vortices in the selfdual Abelian Higgs
theory.  The gauge theory has both Maxwell and Chern-Simons kinetic
terms. The effective Lagrangian consists of the quadratic and linear
in small vortex velocity. We have argued that the valid regime of this
effective Lagrangian is the case where Chern-Simons coefficient is
small. We have also shown that our effective Lagrangian has the
conserved angular momentum consistent with the field theoretic one.
We have studied two vortex dynamics briefly.

In appendix we have shown the first order term induced by the small
background charge and also a harmonic potential well given by the
moment of inertia. We see that the general  consistent dynamics of
vortices contains second and first order terms and a potential. It
would be very interesting to find the detailed dynamics of two vortices 
in this frame. 

The supersymmetric generalization of our Lagrangian would also be very
interesting. In supersymmetric theories~\cite{choonkyu} we expect that 
there may be
many threshold bound states of vortices in large $\kappa$ limit. 
Another intriguing question may be
to see whether or not this is true in our low energy
Lagrangian.

\noindent{\bf Acknowledgement}

Y.K. was supported by the KRF Grant 2001-015-DP0082. 
K.L. was supported in part by KOSEF 1998
Interdisciplinary Research Grant 98-07-02-07-01-5. 
One of us(K.L.) would like to thank Nick Manton for extensive
discussions on the materials in this paper and Organizers of the
M-theory Workshop in the Newton Institute where a part of work is
done.

\newpage

\noindent{\bf Appendix A: Uniform  Background Charge}

We can add the uniform background electric charge density to the Lagrangian,
\beq
{\cal L} _{\rho_e} = -\rho_e A_0 \; .
\eeq
The energy function still gets the BPS bound if we choose the potential to be
\beq
U = \frac{1}{8}(1-f^2+ 2\kappa N)^2 + \frac{1}{2} f^2N^2 - \rho_e N.
\eeq
Interestingly the BPS bound appears with only one sign when one mixes
the Chern-Simons terms and the background charge~\cite{piljin}.  When
both $\kappa$ and $\rho_e$ are very small, we expect the correction of order
$\rho_e v$ and $\kappa \rho_e$. The vacuum structure of this theory is
much more complicated and somewhat similar to those studied in
Ref.~\cite{piljin} (See below in  Table 1).

\begin{tabular}{|c|c|c|c|} \hline
vacuum & $\kappa\rho_{e}$ & $f^{2}$ & $N$ \\ \hline
broken & $\kappa\rho_{e}>-1/8$
& \mbox{ $\displaystyle{
\frac{1+\sqrt{1+8\kappa\rho_{e}}}{2}\stackrel{\rho_{e}\rightarrow 0}{
\longrightarrow}1
}$ }
& \mbox{$\displaystyle{
\frac{2\rho_{e}}{1+\sqrt{1+8\kappa\rho_{e}}}
\stackrel{\rho_{e}\rightarrow 0}{\longrightarrow}0
}$ }\\ \cline{2-4}
& $-1/8< \kappa\rho_{e} \le 0$
& \mbox{$\displaystyle{
\frac{1-\sqrt{1+8\kappa\rho_{e}}}{2}\stackrel{\rho_{e}\rightarrow 0}{
\longrightarrow}0
}$ }
& \mbox{ $\displaystyle{
\frac{2\rho_{e}}{1-\sqrt{1+8\kappa\rho_{e}}}
\stackrel{\rho_{e}\rightarrow 0}{\longrightarrow}-\frac{1}{2\kappa}
}$ }  \\ \cline{2-4}
& $\kappa\rho_{e}=-1/8$
& 1/2 & $2\rho_{e}$ 
\\ \cline{2-4}
& $\kappa\rho_{e}<-1/8$
& not real & not real 
\\ \hline
symmetric & all values & 0 & \mbox{$\displaystyle{
\frac{1}{2\kappa}\left(\frac{2\rho_{e}}{\kappa}-1\right)
\stackrel{\rho_{e}\rightarrow 0}{\longrightarrow}-\frac{1}{2\kappa}
}$ }\\ \hline
\end{tabular}
\begin{center}
{Table 1}
\end{center}

We are interested in selfdual vortices in the
broken phase $\langle f \rangle \approx 1,; \langle N \rangle \simeq
0$ when the small $\kappa$ and $\rho_e$ limit is taken. For
small $\kappa$ and $\rho_e$ limit, their contributions do not
mix each other.  As there is no correction to the mass of vortices,
there is no $\kappa \rho_e$ order correction. The $\rho_e v$
correction in Ref.~\cite{kml} can be written as
\beq
L_{\rho_e v} =  \frac{\pi \rho_e}{2} \sum_r \left\{ 
i \dot{z}_r (\bar{z}_r -H_r) 
 + cc \right\} 
\eeq
whose conserved angular momentum is 
\beq
J_{\rho} = \pi \rho_e \sum_r \left\{ |z_r|^2 + ( z_r H_r + cc ) \right\} .
\eeq

A single vortex feels the uniform magnetic field by the first order
term. This is due to the combination of Lorentz and Magnus forces on
the vortex. When the background charge is very small, the Landau
level energy will be much smaller than the vortex mass, making the low
energy effective action possible. 

As there is BPS bound for only one sign in the energy density, one
suspects that there may be a potential of order $\kappa \rho_e$ for
the moduli space dynamics of antivortices. It remains to be clarified
in future.

\noindent{\bf Appendix B: Harmonic Well Potential}

When the coupling constant for the field theory potential is tiny bit
different from the critical value, the induced potential on the
vortices can be included in the low energy Lagrangian as shown in
Ref.~\cite{shah}.  However, the explicit form of the vortex potential
is not available and so the analysis is more complicated.

On the other hand, one can imagine a different type of potential on
vortices. For example, we can imagine the magnetic flux on two plane
is bundled by an external magnet in third space, like vortices on
superconducting large disc lying between two poles of a magnet. It is
not easy to incorporate such a potential in the field theory
Lagrangian.  However, it is easy to add a `confining' potential on
vortices.

The second order Lagrangian has an obvious rotational symmetry. We
have discussed the corresponding conserved angular momentum
extensively. On the moduli space with the metric by Samols, there
exists a corresponding  Killing vector
\beq
K = \sum_r i(z_r\partial_{z_r} - \bar{z}_r \partial_{\bar{z}_r} ) .
\eeq
This is holomorphic Killing vector. In  the
discussion of 1/4 BPS magnetic monopole dynamics~\cite{piljin,tong},
the additional potential was given by the quadratic norm of the
triholomorphic Killing vectors for the internal symmetry. Similarly,
we find the quadratic norm of the angular momentum Killing vector as a
potential 
\beq
U_{\rm eff} = \frac{\lambda}{2} g_{rs} z_r \bar{z}_s
\eeq
with a small positive coefficient $\lambda$.  This potential leads to
the harmonic potential well to the vortices. 

The quadratic norm of the angular momentum Killing vector is of course
the moment of inertia we know well. Thus the above potential is
proportional to the inertia. The inertia can be split into one for the
center of mass motion and another for the relative motion. Thus, we
can have one potential for the center of mass motion and another
potential for the relative motion with different coefficient.

When $\lambda$ is of order $v^2$, we can include this in the low
energy Lagrangian. 
When vortices are close to each other, the configuration of them
is very complicated but roughly one can see
that vortices are incompressible particles on two dimensions. The
reason is that $\int d^2x (1-f^2)/(4\pi)=n $ measures the vortex core
area and is increasing with vortex number. Our potential is a harmonic
well and so the low energy dynamics of $L_{v^2} +U_{\rm eff}$ describes
particles of finite size hard core and harmonic well. All interactions
are short ranged. Of course one can add the first order Lagrangians
too. As the angular momentum Killing vector is holomorphic, the
supersymmetry can be extended to include the potential.

\end{document}